\documentclass[showpacs,amsmath,
preprintnumbers,
twocolumn,
superscriptaddress,
aps,prb]{revtex4}
\usepackage{graphicx}
\usepackage{dcolumn}

\begin{document}

\title{Antisite effect on ferromagnetism in (Ga,Mn)As}
\author{R. C. Myers}
\affiliation{Center for Spintronics and Quantum Computation, 
University of California, Santa Barbara, CA 93106.}
\author{B. L. Sheu}
\affiliation{Department of Physics and Materials Research Institute, 
The Pennsylvania State University, University Park, PA 16802.}
\author{A. W. Jackson}
\author{A. C. Gossard}
\affiliation{Center for Spintronics and Quantum Computation, 
University of California, Santa Barbara, CA 93106.}
\author{P. Schiffer}
\author{N. Samarth}
\affiliation{Department of Physics and Materials Research Institute, 
The Pennsylvania State University, University Park, PA 16802.}
\author{D. D. Awschalom}
\affiliation{Center for Spintronics and Quantum Computation, 
University of California, Santa Barbara, CA 93106.}

\date{\today}

\begin{abstract}
We study the Curie temperature and hole density of (Ga,Mn)As while 
systematically varying the As-antisite density. Hole compensation by 
As-antisites limits the Curie temperature and can completely quench 
long-range ferromagnetic order in the low doping regime of 1-2{\%} Mn. 
Samples are grown by molecular beam epitaxy without substrate rotation in 
order to smoothly vary the As to Ga flux ratio across a single wafer. This 
technique allows for a systematic study of the effect of As stoichiometry on 
the structural, electronic, and magnetic properties of (Ga,Mn)As. For 
concentrations less than 1.5{\%} Mn, a strong deviation  from $T_{\mathrm{C}} \propto p^{0.33}$ is observed. Our results emphasize that proper control of 
As-antisite compensation is critical for controlling the Curie 
temperatures in (Ga,Mn)As at the low doping limit.
\end{abstract} 

\maketitle

\section{Introduction}
\label{Sec:intro}

In (Ga,Mn)As, currently the most studied dilute magnetic semiconductor (DMS), a long-range 
ferromagnetic interaction between dilute Mn$^{2 + }$ spins is mediated by valence band (or defect band) hole 
spins.\cite{Ohno:1998,Dietl:2000,Jungwirth:2006} Most 
experimental work in this system has focused on achieving maximal Curie 
temperatures ($T_{\mathrm{C}}$'s) in a technological drive to bring ferromagnetism in 
(Ga,Mn)As to more practical temperatures.\cite{Macdonald:2005} In this 
vein, it was determined that a crucial limiting parameter of the $T_{\mathrm{C}}$ of 
as-grown (Ga,Mn)As is the incorporation of hole compensating crystalline 
defects, which are unintentionally formed due to the low growth temperature 
($\sim $250\r{ }C) and low solubility of Mn in GaAs. One such defect, the Mn 
interstitial (Mn$_{\mathrm{i}}$), is considered to be a double donor compensating two 
holes each.\cite{Erwin:2002} These defects have shown a surprisingly high 
diffusivity allowing for their removal by using post-growth annealing at relatively low 
temperatures ($\sim $200\r{ }C) that drive them to the 
surface.\cite{Edmonds:2004} Consequently, the three dimensional hole density 
($p$) and $T_{\mathrm{C}}$ after annealing can be drastically 
increased.\cite{Ku:2003} Another well-studied defect in low 
temperature (LT) grown GaAs is excess arsenic. For substrate temperatures 
less than 300\r{ }C, excess arsenic is incorporated by occupying a Ga-site, 
called an antisite defect (As$_{\mathrm{Ga}}$) which - like the Mn$_\mathrm{i}$ defect is also a double donor.\cite{Liu:1995} The 
density of these defects can reach 10$^{20}$ cm$^{ - 3}$, $\sim $1{\%} of 
Ga-sites, which is the same order of magnitude as typical Mn doping 
levels.\cite{Missous:1994} By post-growth annealing at temperatures above 
500\r{ }C, As$_{\mathrm{Ga}}$ can be removed by the formation of metallic As 
precipitates,\cite{Bliss:1992} however at these temperatures Mn also 
precipitates out of the (Ga,Mn)As solid solution as MnAs 
nano-particles.\cite{Boeck:1996}

Most experimental work on (Ga,Mn)As has been aimed at maximizing the $T_{\mathrm{C}}$  by 
controlling the compensating defects in the high doping regime, i.e. Mn 
$\sim $5-9{\%}. Such studies mainly focus on the reduction of Mn$_{\mathrm{i}}$ defects 
through post-growth annealing while loosely treating the problem of As 
non-stoichiometry. As$_{\mathrm{Ga}}$ defects are usually ignored in the high doping regime 
since 
Mn$_{\mathrm{i}}$ defects are the main source of carrier compensation, and their removal has 
led to dramatic increases in $p$ and $T_{\mathrm{C}}$.\cite{Ku:2003} On the 
other hand, hole compensation due to As$_{\mathrm{Ga}}$ could be as large as 
10$^{20}$ cm$^{ - 3}$, which would shift $T_{\mathrm{C}}$ in this Mn-doping range by 
up to 10 K.

In contrast, near the ferromagnetic threshold of $\sim $ 1{\%} Mn, compensation 
due to As$_{\mathrm{Ga}}$ could dramatically alter the $T_{\mathrm{C}}$'s since here the 
defect and dopant densities are roughly equal. Additionally, the 
concentration of Mn necessary for the onset of ferromagnetism should be a 
strong function of carrier compensation since the ferromagnetism depends 
strongly on $p$; \textit{in the low doping limit ferromagnetism is likely extrinsically limited by hole compensating defects}. We also note that theoretical work predicts a change in the 
mechanism of ferromagnetism in the low doping limit, near the 
insulator-metal transition,\cite{Erwin:2002,Yang:2003} compared 
with the metallic limit at which $T_{\mathrm{C}} \propto p^{0.33}$ as 
predicted by the Zener-model.\cite{Dietl:2000}

The effect of excess As on the lattice constant of (Ga,Mn)As was treated by 
Schott \textit{et al.},\cite{Schott:2001} who used a different As-flux than another 
study\cite{Ohno:1996} and noted a change in the extrapolated lattice 
constant of zinc blende MnAs due to As$_{\mathrm{Ga}}$ incorporation. They 
followed-up their study by using 2 different As:Ga beam flux ratios, differing by 
500{\%} and noted large changes in the lattice constant of (Ga,Mn)As over a broad 
range of Mn-doping and substrate temperatures.\cite{Schott:2003} These 
results were elaborated upon by Sadowski and Domagala, who measured lattice 
constant variation in Ga$_{0.96}$Mn$_{0.04}$As grown using 3 different As:Ga 
flux ratios and noted an interplay between Mn$_{\mathrm{i}}$ and As$_{\mathrm{Ga}}$ incorporation 
for this Mn doping regime.\cite{Sadowski:2004} Campion \textit{et al.} reported an 
improvement in structural and magnetic properties of (Ga,Mn)As films grown 
using As$_{2}$ instead of As$_{4}$.\cite{Campion:2003} Their paper, 
however, does not describe the fluxes used for each As-species nor how the 
effective III-V ratio varies between these cases. The same group has 
reported impressively high $T_{\mathrm{C}}$'s over a broad range of Mn-doping levels, 
mentioning the importance of minimizing the As:Ga flux ratio although no data on 
the variation of this parameter were provided.\cite{Foxon:2005} Avrutin \textit{et al.} 
carried out the first study that measured both the structural, electrical, 
and magnetic properties of (Ga,Mn)As using 2 different As:Ga flux
ratios.\cite{Avrutin:2005} They noted an optimization of the hole density, 
conductivity, and $T_{\mathrm{C}}$ for the low As:Ga flux ratio, where fewer As$_{\mathrm{Ga}}$ 
should be incorporated.

In this paper, we present the first systematic investigation of the effect 
of As$_{\mathrm{Ga}}$ incorporation on the structural, electronic, and magnetic 
properties of (Ga,Mn)As in the low Mn-doping regime. The As:Ga flux ratio is smoothly varied within individual 
samples by using molecular beam epitaxy (MBE) without substrate rotation. 
The geometry of the MBE chamber inherently provides a gradient in the As:Ga 
flux ratio across the substrate of $\pm $50{\%}, while maintaining a roughly 
uniform Mn flux. In order to avoid formation of Mn$_{\mathrm{i}}$ and to achieve a 
high sensitivity to carrier compensation, we chose Mn doping concentrations 
of 1 - 2{\%}, in the region of the onset of ferromagnetism. At Mn 
concentrations less than 1.5{\%}, Mn$_{\mathrm{i}}$ has a larger formation energy 
than Mn$_{\mathrm{Ga}}$.\cite{Jungwirth:2006} Thus no interstitial Mn will form and 
all hole compensation should originate from As$_{\mathrm{Ga}}$. Film morphology is 
measured \textit{in-situ} by reflection high energy electron diffraction (RHEED) and 
\textit{ex-situ} by atomic force microscopy (AFM). Out-of-plane epitaxial strain is measured 
by high-resolution x-ray diffraction (HRXRD). Electronic properties at room 
temperature are measured by Hall effect in the Van der Pauw and Hall-bar 
geometry, and magnetic properties are measured by SQUID magnetometry (Quantum Design MPMS).

The epitaxial strain due to the incorporation of As$_{\mathrm{Ga}}$ can be controlled 
by the As:Ga flux ratio, as previously 
reported.\cite{Liu:1995,Sadowski:2004} More interestingly, 
the electronic and magnetic properties reveal a strong sensitivity to 
As$_{\mathrm{Ga}}$ compensation with conductivities and hole densities ($p$) spanning 
over two orders of magnitude leading to a transition from paramagnetism to 
ferromagnetism for constant Mn densities (Sec. \ref{Sec:rotated} and \ref{Sec:electronic}). Hole compensation due to As$_{\mathrm{Ga}}$ 
reduces $p$ and therefore $T_{\mathrm{C}}$. The As:Ga flux ratio is also observed to 
significantly alter film morphology (Sec. \ref{Sec:structural}). In the As-rich condition, films attain 
maximum smoothness, show a high degree of As$_{\mathrm{Ga}}$ related carrier 
compensation, and exhibit reduced or suppressed Curie temperatures. For As:Ga flux ratios at which the $p$ and therefore $T_{\mathrm{C}}$ are 
maximized (stoichiometric condition, As$_{\mathrm{Ga}} \sim  0$), the films remain 
two-dimensional (2D) but are rougher than the As-rich material. In the 
Ga-rich condition (small As:Ga flux ratios), films become three-dimensional (3D) with 
the accumulation of excess Ga on the surface leading to an order of 
magnitude increase in film roughness. At the minimum As:Ga flux ratio
investigated, the roughest surfaces are observed with hemispherical 
Ga-droplets. The group-III enriched surface suppresses Mn incorporation 
during growth, leading to lower $p$ and $T_{\mathrm{C }}$ values than in the As-rich 
condition. In the low Mn-doped regime, where As-compensation limits hole 
density and ferromagnetic ordering, a strong deviation from the usual 
$T_{\mathrm{C}}  \propto p^{0.33}$ behavior is observed with $T_{\mathrm{C}} \propto 
p^{0.09}$ for $x \le $ 1.5{\%}. Our results emphasize that proper 
control of hole compensation due to As$_{\mathrm{Ga }}$ is critical to maximizing the value of $T_{\mathrm{C}}$'s in the low Mn-doping regime. The combinatorial MBE growth 
approach used here could be applied to more complex LT grown GaAs-based 
heterostructures in which the As:Ga flux ratio is a strong control parameter.

\section{Molecular Beam Epitaxy (MBE) growth}
\label{Sec:MBE}

All samples are grown on 2-inch diameter GaAs (001) substrates in a Varian 
Gen-II MBE system manufactured by Veeco Instruments, Inc. with a source to substrate distance of 5.4-inches. The substrate 
temperature is monitored by absorption band edge spectroscopy (ABES), which 
measures the substrate temperature in real-time through white-light 
transmission spectroscopy. The temperature stability during LT (Ga,Mn)As and 
GaAs growths is typically $\pm5$\r{ }C . The growth rate of GaAs at the 
center of the wafer is 0.7 ML/sec as calibrated by RHEED intensity 
oscillations of the specular spot at 580\r{ }C. The Mn doping density is 
inferred from RHEED intensity oscillation measurements\cite{Schippan:2000} 
of the MnAs growth rate performed at 240\r{ }C. Measurements of the Mn density by secondary ion mass spectroscopy (SIMS) indicate that the RHEED calibration underestimates the density by up to 0.4{\%} Mn, which provides a rough error estimate for the quoted Mn concentrations. We use As$_{2}$ for all 
samples presented here, employing a valved cracker/sublimator. The beam 
equivalent pressure (BEP) of As$_{2}$ and Ga are measured with a nude ion 
gauge at the center position of the substrate just prior to growth. The 
ratio of As$_{2}$ and Ga BEPs, defined as $As:Ga$, is proportional to the 
ratio of the atomic fluxes of As and Ga delivered to the growing surface. 
Unfortunately, a reliable quantitative conversion from BEP to atomic flux is 
difficult due to the uncertainties in the ionization efficiencies of each 
beam species.\cite{Farrow:1995} A rough estimate gives that the atomic flux ratio is 2.5 times 
smaller than the BEP ratios, where we assume that only As$_{2}$ are present 
with a temperature equal to that of the cracking zone. 

The following procedure is used for both non-rotated and rotated growths 
prior to deposition of the low temperature (LT) films. The wafer is first 
heated to 635\r{ }C with $As:Ga \sim 35$ for oxide desorption, monitored by 
RHEED, and subsequently cooled to 585\r{ }C. While rotating at 10 RPM, a 
300-nm GaAs buffer layer is grown using 5-s growth interrupts for every 15 
nm of deposition resulting in a streaky 2 $\times $ 4 surface reconstruction 
RHEED pattern. While cooling the substrate to 250\r{ }C, the As-valve 
setting is reduced at 400\r{ }C for the desired $As:Ga$ value for LT growth. 
Prior to LT growth, a streaky 4 $\times $ 4 surface reconstruction is 
observed in the RHEED pattern. For rotated growths, once the substrate 
stabilizes at 250\r{ }C the LT growth begins. For non-rotated growths, the 
substrate rotation is stopped at 400\r{ }C and the wafer [110] crystal axis (perpendicular to the wafer major flat) 
is aligned along the $As:Ga$ gradient direction (y-axis). This can be 
reproducibly accomplished noting the radial angle between the RHEED e-beam 
and substrate, fixed by the MBE chamber geometry, and finding a high 
symmetry crystal axis from the surface reconstruction pattern. All LT films 
are 100-nm thick at the center wafer position. After completion of the LT 
layer, they are cooled in a consistent manner to room temperature.

\section{Rotated (Ga,Mn)As: carrier compensation and quenched 
ferromagnetism}
\label{Sec:rotated}

\begin{figure}\includegraphics{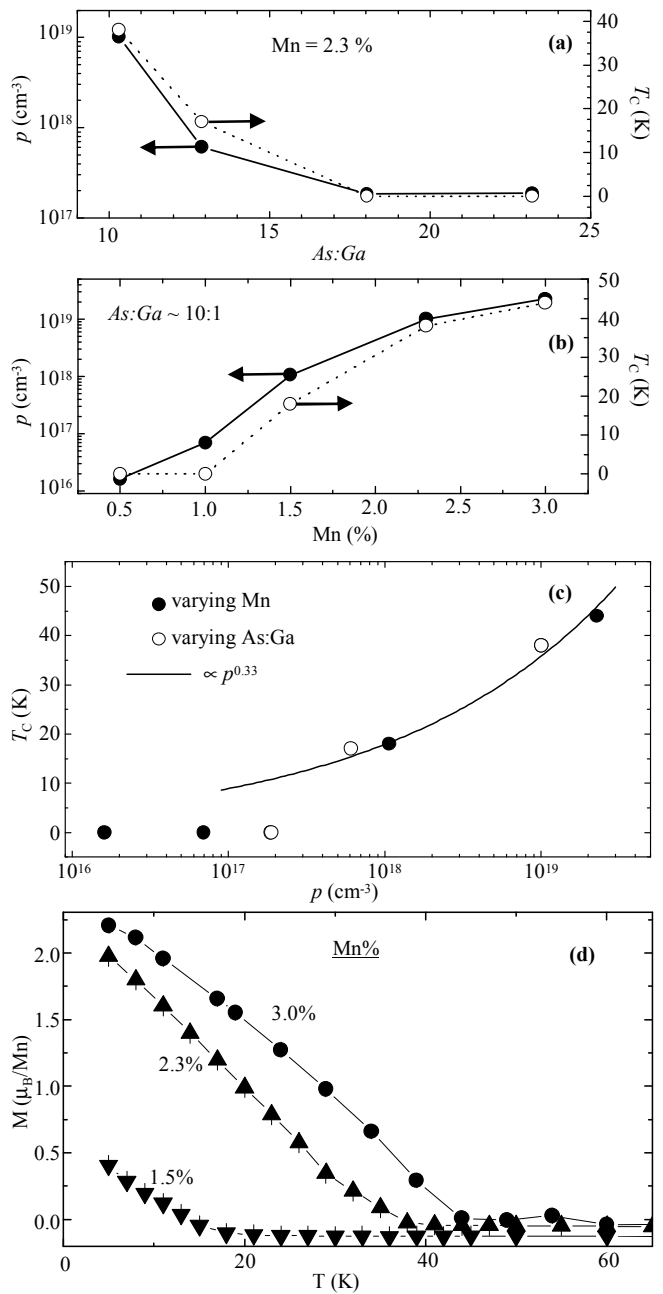}\caption{\label{fig1}
    Quenched ferromagnetism by As-compensation. Data from (Ga,Mn)As films 
		grown at 250\r{ }C using substrate rotation. (a) Samples were grown using 
		a constant growth rate and Mn concentration but varying $As:Ga$. Room 
		temperature hole density ($p$) is plotted on the left vertical axis as a 
		function of $As:Ga$. The corresponding Curie temperature ($T_{\mathrm{C}}$)
		is plotted on the right vertical axis. (b) Similar data plotted for samples 
		grown with a constant $As:Ga$, but varying Mn concentration. (c) Data in 
		(a) and (b) are replotted with $T_{\mathrm{C}}$ as a function of $p$. A fit to $T_{\mathrm{C}} 
		 \propto p^{0.33}$ is shown for the ferromagnetic samples. (d) 
		Magnetization versus temperature scans (field-warmed with B = 50 Oe) used to 
		extract $T_{\mathrm{C}}$ shown in (b).}\end{figure}

We first consider the case of rotated growth, examining the effect of Mn 
concentration and $As:Ga$ on the electronic and magnetic properties. Four 
samples are grown with the same Mn concentration (2.3{\%}), but with various 
$As:Ga$. Hole density data are from samples measured in the Van der 
Pauw geometry in a magnetic field up to 0.2 T. Figure \ref{fig1}a plots $p$ and $T_{\mathrm{C}}$
of these samples as a function of $As:Ga$. Values of $T_{\mathrm{C}}$ are obtained from 
magnetization versus temperature scans as shown in Fig. \ref{fig1}d selecting the point of maximum second derivative in M(T). The scans are performed during a field warm (50 Oe) after a field cool (1 T). The hole density 
decreases exponentially with the $As:Ga$, indicating that excess As in the form of 
As$_{\mathrm{Ga}}$ donors are compensating Mn$_{\mathrm{Ga}}$ acceptors. Correspondingly, 
$T_{\mathrm{C}}$ decreases with $p$, and for $As:Ga \ge 18$ ferromagnetism is 
suppressed. This strong compensation effect is due to the fact that the 
As$_{\mathrm{Ga}}$ donor and Mn$_{\mathrm{Ga}}$ acceptor densities are similar in magnitude in 
this low Mn-doping regime, Mn $<$ 2 {\%}. 

Using $As:Ga = 10.3$, for which the largest $p$ and $T_{\mathrm{C}}$ are obtained (Fig. 
\ref{fig1}a), a series of four additional samples are grown with varying Mn 
concentration. For this series, variation in $p$ and $T_{\mathrm{C}}$ occurs due to the 
varying concentration of Mn$_{\mathrm{Ga}}$, Fig. \ref{fig1}b. Clearly the hole density does not 
increase linearly with Mn concentration, which is indicative of a 
carrier compensation threshold. Based on this observation, a more careful 
optimization of $As:Ga$ is necessary, as discussed in the next sections.

Figure \ref{fig1}c plots $T_{\mathrm{C}}$ as a function of $p$, summarizing data from Fig. \ref{fig1} a 
and b. We observe that varying the concentration of compensating As$_{\mathrm{Ga}}$ 
donors (by varying $As:Ga$) has a qualitatively identical effect as varying 
the concentration of magnetic Mn$_{\mathrm{Ga}}$ acceptors with respect to the 
carrier density and magnetic transition temperature. A fit of $T_{\mathrm{C}} 
\propto p^{0.33}$ to the ferromagnetic samples in Fig. \ref{fig1}c, Mn doping $ 
\ge $ 1.5{\%}, reveals the usual functional dependence.\cite{Macdonald:2005} 

\section{Non-rotated (Ga,Mn)As: combinatorial variation of $As:Ga$}
\label{Sec:nonrotated}

In order to systematically study the effect of the excess As on the 
electronic and magnetic properties of (Ga,Mn)As, we use a method of non-rotated 
growth in which the geometry of the MBE system provides a continuous 
variation in $As:Ga$ across the wafer.\cite{Jackson:1} Thus the 
advantage of a combinatorial approach is employed in which a single 
non-rotated growth provides the equivalent of twenty rotated growths of 
various $As:Ga$. The phase space of $As:Ga$ and Mn concentration can be 
mapped-out in high-resolution while performing a much smaller number of 
growths than otherwise required.

\begin{figure}\includegraphics{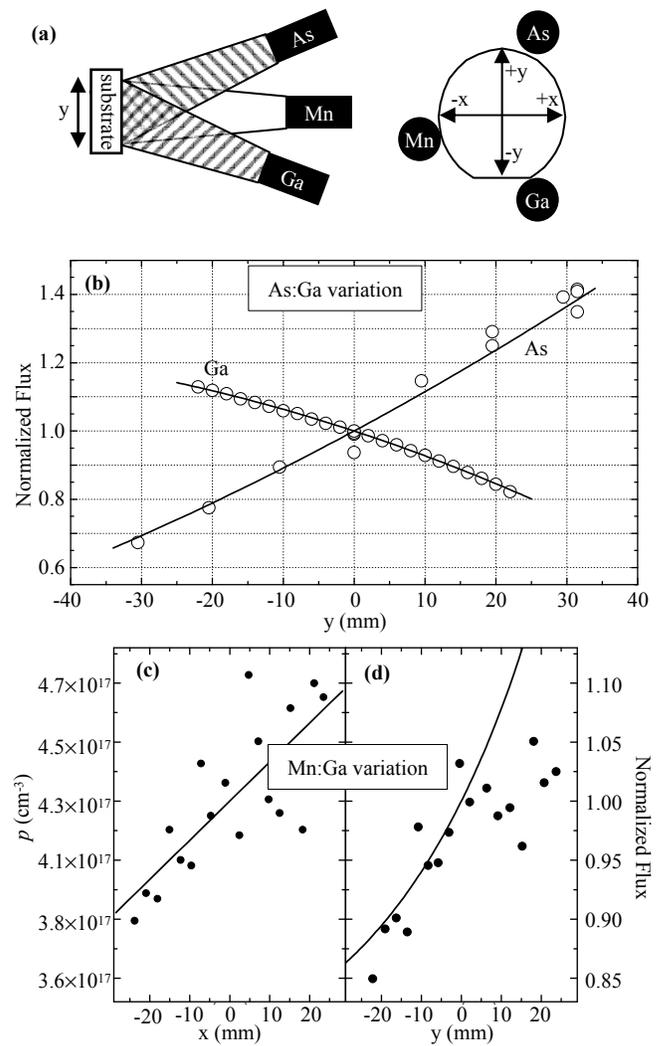}\caption{\label{fig2}
    Geometric flux gradients in the MBE system. $As:Ga$ is 
		continuously varied and optimized in a single non-rotated growth. (a) 
		Deposition geometry (not to scale) in the MBE system. The relative positions of the As, Ga, 
		and Mn Knudsen cells with respect to the substrate are shown from side and 
		front views. (b) The flux gradient of Ga and As molecular beams along the 
		y-axis of the wafer normalized to the center flux value. Measured values 
		(data points) and fits to second order polynomials (lines) are shown. 
		(c)-(d) The Mn:Ga flux variation along the x-axis (c) and y-axis (d) of the 
		wafer measured from hole density ($p$) variation in non-rotated samples grown 
		at 400 C, where As-flux does not affect carrier density. The line in (d) is 
		the inverse of the Ga-flux variation, shown as a line in (b), which accounts 
		for the variation in the Mn:Ga ratio along the y-axis.}\end{figure}

Figure \ref{fig2}a schematically shows the position of the Ga Knudsen cell and the As 
valved-cracker/sublimator relative to the substrate position. The angle 
between the Ga and As$_{2}$ molecular beams provides a continuous variation 
in $As:Ga$ across the y-axis of the substrate, as defined in the 
schematic. The variation of the As flux is measured by depositing a thick As 
film on a quartz wafer, covered by a shadow mask along the y-axis, with the 
substrate stabilized below room temperature. We have assumed that the 
deposition rate of As under these conditions is proportional to the As flux. 
A profilometer is used to measure the thickness of the As film along the 
y-axis. The normalized flux variation of As, plotted in Fig. \ref{fig2}b, is 
calculated by normalizing the thickness along y by the thickness at y = 0. 
The variation of the Ga-flux is measured using an optical measurement of a 
distributed Bragg reflector (DBR) and cavity structure grown in 2 separate 
steps. First, a 10 period GaAs/AlGaAs DBR is grown on a rotated 2-inch 
substrate in a larger, more uniform system (Veeco Gen III) with a source to 
substrate distance of 11-inches. This results in an extremely uniform DBR 
which is As-capped and transferred into the Gen II system. The As-cap is 
desorbed and a GaAs optical cavity layer is grown on the sample without 
substrate rotation. The reflectance spectrum of the sample is measured 
\textit{ex-situ} at different points across the wafer and is fit to an optical model of the 
reflectance. Because the 10 period DBR was almost completely uniform, the 
thickness of the non-uniform GaAs layer grown in the Gen II could be 
accurately and unambiguously determined by fitting the measured optical 
spectrum to a computer model of the structure. Since the deposition is 
performed at 585\r{ }C, the growth rate is independent of the As 
overpressure, being limited by the group-III flux only. The normalized Ga 
rate, proportional to the Ga-flux, along the y-axis is plotted in Fig. \ref{fig2}b. 
Both the As and Ga flux variation are well fit by a second order polynomial 
(lines), which are used to convert y-axis position for non-rotated growths 
to $As:Ga$.

Next we consider the Mn flux variation across the substrate. Since the Mn 
Knudsen cell is roughly perpendicular to the y-axis, variation along this 
direction is expected to be small. In order to measure the Mn variation 
across the substrate, a 1-$\mu $m thick GaAs:Mn layer with Mn $\sim $ 
4$\times $10$^{17}$ cm$^{ - 3}$ is deposited at 400\r{ }C with an 
As$_{2}$ overpressure. At these conditions Mn incorporation is fully 
substitutional and excess As incorporation is negligible ($<$ 1 $\times $ 
10$^{16}$ cm$^{ - 3})$, thus the hole density is proportional to the Mn:Ga 
flux ratio.\cite{Poggio:2005} The hole density variation across the x and y 
axes due to Mn:Ga variation is plotted in Fig. \ref{fig2} c and d, respectively. 
Along either axis, the Mn:Ga ratio varies by 5-10{\%}, which would correspond to 
variation in the Mn concentrations of 0.05 to 0.1{\%}. As will be clear (Sec. \ref{Sec:electronic}), this variation in Mn concentration is negligible with 
respect to the As compensation effect under investigation.

\section{Film morphology and strain versus $As:Ga$}
\label{Sec:structural}

\begin{figure}\includegraphics{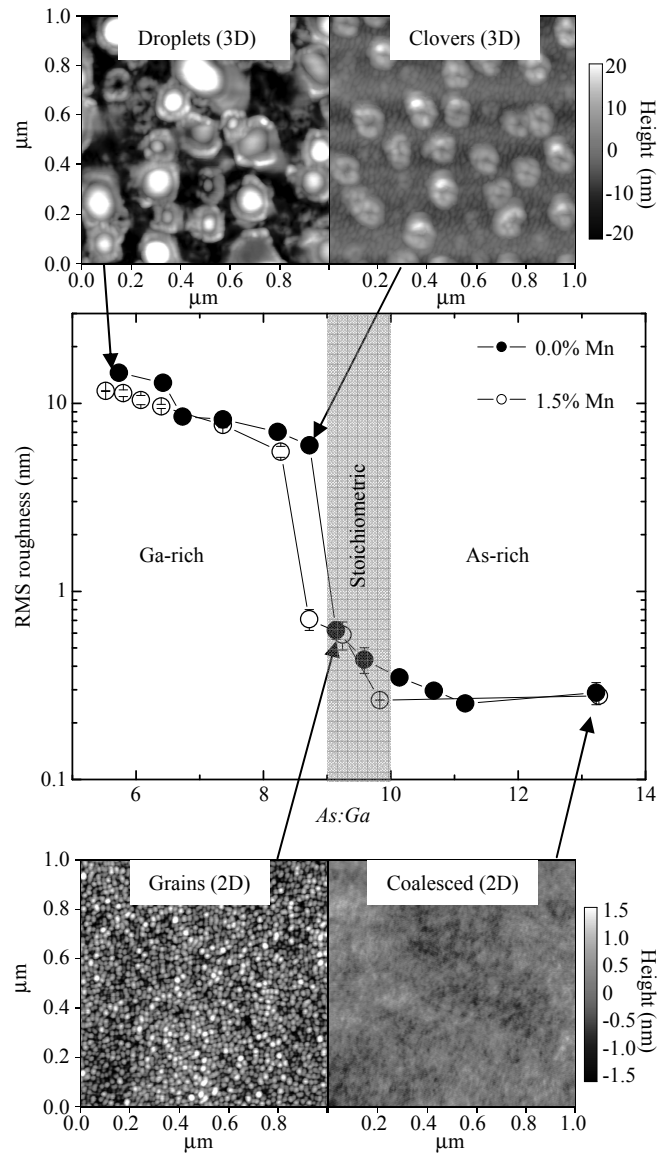}\caption{\label{fig3}
    Morphology of GaAs and (Ga,Mn)As films as a function of $As:Ga$ 
		(non-rotated growths). AFM micrographs (1$\times $1 $\mu $m$^{2}$ scans) are 
		plotted for a GaAs growth, but are qualitatively identical to (Ga,Mn)As films 
		in the same wafer position. The RMS roughness measured from such AFM scans 
		is plotted as a function of $As:Ga$ (along y-axis of wafer) for both 
		a GaAs and (Ga,Mn)As sample. The stoichiometric region is shaded grey, where 
		excess As is minimized, two-dimensional growth (2D) is maintained, however, 
		some granular structure is observed.
    }\end{figure}

To study the effect of $As:Ga$ on the film morphology, the RMS roughness of 
non-rotated samples is measured across the y-axis by AFM. We describe 
results from two non-rotated films, without Mn doping and with 1.5{\%} Mn. 
Both 5$\times $5-$\mu $m$^{2}$ and 1$\times $1-$\mu $m$^{2}$ area scans are 
measured every 3 to 4 mm along the y-axis of both wafers. The RMS roughness 
from these scans is plotted as a function of $As:Ga$ (converted from 
y-position from Fig. \ref{fig2}b calibration) and representative AFM images of each 
region are shown in Fig. \ref{fig3}.

In the As-rich region from $11 < As:Ga < 14$, AFM images reveal fully 
coalesced films with RMS roughness of 0.2 nm or less indicative of a 2D growth mode. RHEED observations of 
such films, from rotated growths at the same $As:Ga$ condition, show sharp and 
streaky 2D reconstruction. In the As-rich to stoichiometric region from $9 < 
As:Ga < 11$, films show a nanoscale-granular structure. For decreasing $As:Ga$, 
this granular structure becomes more distinct and the film roughness 
increases, however it remains below 1 nm RMS. RHEED observations for films 
grown in this $As:Ga$ range show a spotty reconstruction pattern indicative of 
a rough, but 2D surface, which we have found to be characteristic of films 
near the stoichiometric condition. In this region, the hole density of 
Mn-doped films reaches a maximum indicating the minimum of compensating 
As$_{\mathrm{Ga}}$ defects (Sec. \ref{Sec:electronic}).

Just below $As:Ga \sim 9$, the RMS film roughness increases by a factor of 
ten as $As:Ga$ transitions from the stoichiometric condition to 
Ga-rich. AFM images of this region show the formation of 10-nm tall 
clover-shaped features 60-nm in diameter with a crater in the center. The 
RHEED pattern of these films shows additional spots indicative of 3D 
features on the surface. For $As:Ga < 9$, the surface roughness continues to 
increase with the appearance of larger droplet-like features. At the lowest 
$As:Ga$ ratio, these hemispherical droplets are 50-nm tall and 240-nm in 
diameter. The large change in roughness across the non-rotated wafers is 
observable by eye under intense illumination, where the surface transitions 
from mirror finish at the As-rich region to a hazy surface in the Ga-rich 
region.

As observed in the roughness plot of Fig. \ref{fig3}, it appears that Mn-doping 
suppresses the onset of roughening toward smaller $As:Ga$. In contrast, since 
Mn-doping increases the effective group-III flux, one would expect a shift 
in the onset of roughness to higher $As:Ga$. This smoothing may be due to a 
surfactant effect of Mn on the surface. As discussed in Sec. \ref{Sec:electronic}, the 
highest hole densities are observed in (Ga,Mn)As grown with $As:Ga$ just before 
the transition from 2D to 3D surface, indicating that the growth kinetics at 
this stoichiometric condition do not significantly increase the density of 
hole compensating Mn$_{\mathrm{i}}$ donors.

\begin{figure}\includegraphics{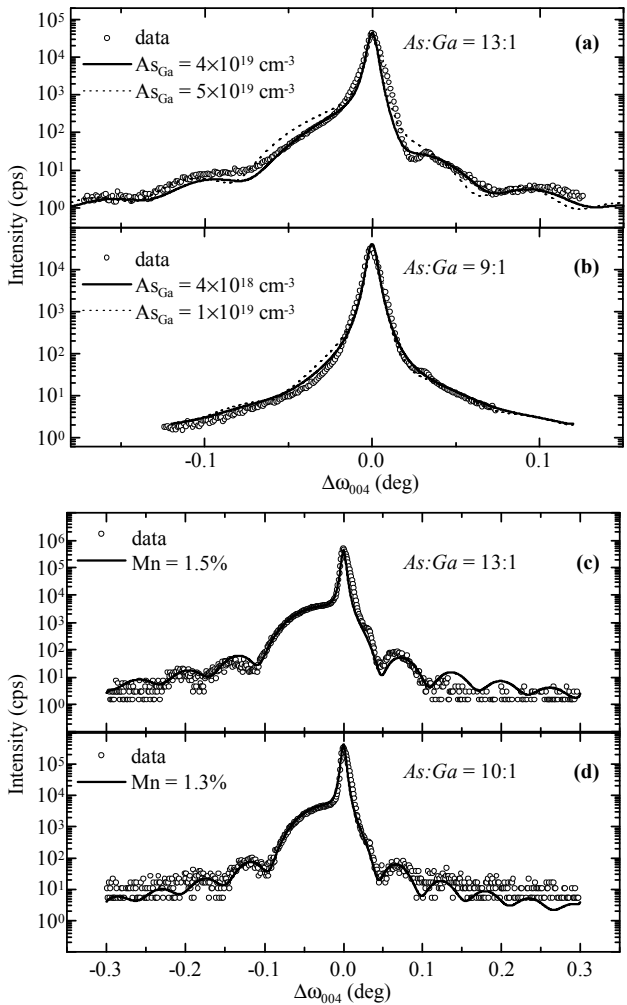}\caption{\label{fig4}
    Film strain due to As non-stoichiometry and Mn incorporation in GaAs 
		and (Ga,Mn)As non-rotated films measured by $\omega $-2$\theta $ HRXRD scans 
		near the (004) peak. (a)-(b) Data from a GaAs film at the As-rich wafer 
		position (a) and stoichiometric position (b). Data (points) and dynamical 
		simulations (lines) are plotted for various As$_{\mathrm{Ga}}$ densities. (c)-(d) 
		Data from a (Ga,Mn)As film at the As-rich wafer position (c) and stoichiometric 
		position (d). Data (points) and dynamical simulations (lines) are plotted 
		using different Mn densities.}\end{figure}

HRXRD scans are obtained using a triple-axis (Phillips X'Pert MRD Pro) thin 
film diffractometer. We measure $\omega $-2$\theta $ scans near the (004) 
Bragg peak of the (001) GaAs substrate. The diffraction patterns obtained 
for the non-rotated LT GaAs film at the position of highest As-compensation 
($As:Ga = 13:1$) and at the stoichiometric condition ($As:Ga = 9:1$) are plotted 
in Fig. \ref{fig4} a and b, respectively. In the As-rich position, the epilayer peak 
appears as a shoulder at lower angle, reflecting the increase in the 
out-of-plane lattice constant, due to As$_{\mathrm{Ga}}$ incorporation. We also note 
the thickness fringes indicative of the high quality of the film surface and 
interface. A dynamical HRXRD program is used to simulate the 
data\cite{Brandt:2002} assuming that the lattice constant of (Ga$_{1 - 
z}$,As$_{z})$As, a = 5.654 + 1.356z.\cite{Liu:1995} Best fit 
simulations in Fig. \ref{fig4}a give a film thickness 90 nm and As$_{\mathrm{Ga}}$ = 4$\times 
$10$^{19}$ cm$^{ - 3}$. The additional fit with As$_{\mathrm{Ga}}$ = 5$\times 
$10$^{19}$ cm$^{ - 3}$ shows the error in the As$_{Ga }$ estimate is about 
1$\times $10$^{19}$ cm$^{ - 3}$. The thickness of the epilayer is 10-15{\%} 
thinner at the top edge than at the center of the wafer due to the smaller 
Ga-flux at this position, as shown in Fig. \ref{fig2}b. At the stoichiometric 
position the epilayer shoulder becomes small, only visible as a slight 
asymmetry on the left side of the substrate diffraction peak in Fig. \ref{fig4}b. 
Simulations using a film thickness of 100 nm yield As$_{\mathrm{Ga}}$ = 4$\times 
$10$^{18}$ cm$^{ - 3}$, which is smaller than the uncertainty of the fits. A 
more precise estimate of the As$_{\mathrm{Ga}}$ density is found from the hole 
density variation, discussed in Sec. \ref{Sec:electronic}. Thickness fringes 
disappear since the lattice constant of the epilayer is almost 
indistinguishable from that of the substrate.

Figure \ref{fig4} c and d plot the diffraction patterns from a non-rotated (Ga,Mn)As 
film with 1.5{\%} Mn from the As-rich and stoichiometric positions, 
respectively. For these layers, strain due to Mn incorporation dominates the 
epilayer lattice constant obscuring the As$_{\mathrm{Ga}}$ effect. Simulations are 
performed assuming that the lattice constant of (Ga$_{1 - x}$,Mn$_{x})$As, a 
= 5.654 + 0.476x.\cite{Schott:2001} The fits indicate a film thickness 
of 85 nm and 95 nm, and Mn concentration of 1.5{\%} and 1.3{\%} at the 
As-rich and stoichiometric positions, respectively. The variation in Mn 
concentration is not explained by the Mn:Ga ratio variation along y, Fig. 
\ref{fig2}d, which would predict a decrease in Mn concentration of up to 0.1{\%}. The 
additional strain reduction may be due to the minimization of As$_{\mathrm{Ga}}$ 
defects at the stoichiometric position, but the uncertainty of the HRXRD 
simulations prevents a more quantitative analysis.

\section{Carrier density and Curie temperature versus $As:Ga$}
\label{Sec:electronic}

\subsection{Hole density gradient}
\label{Subsec:pgradient}

\begin{figure}\includegraphics{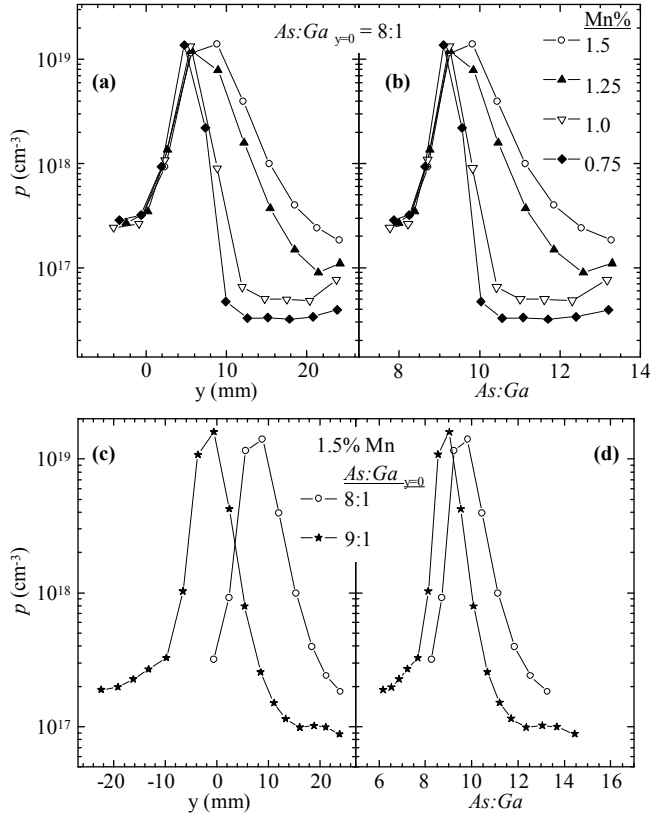}\caption{\label{fig5}
    The As-compensation gradient in non-rotated (Ga,Mn)As films. The room 
		temperature hole density ($p$) variation along films grown with different Mn 
		concentrations, (a) as a function of y-position, and (b) as a function of 
		$As:Ga$. (c)-(d) Similar data for samples with the same Mn 
		concentrations but different $As:Ga$ at y = 0. Lines in each plot 
		connect data points from the same wafer.}\end{figure}

Figure \ref{fig5} a and b show the variation in hole density for a series of 
non-rotated (Ga,Mn)As films of various Mn concentrations (a) along the y-axis, 
and (b) with y converted to $As:Ga$, using the Fig. \ref{fig2}b calibration. All the 
hole density data in the figures are from room temperature measurements of 
samples prepared in the Van der Pauw geometry (3$\times $3 mm$^{2})$ in a 
magnetic field up to 0.2 T. Selective measurements of the hole density of 
patterned Hall-bars in magnetic fields up to 2 T agree with the Van der Pauw 
measurements to within 50{\%}. Additional high field measurements at 7-9 T deviate from the low field data by up to 50 {\%}.

The hole density varies by over two orders of magnitude across each wafer 
showing a maximum value near y $\sim $ 6 mm corresponding to $As:Ga \sim 9$. The results are explained by a decreasing number of As$_{\mathrm{Ga}}$ donors that 
compensate Mn$_{\mathrm{Ga}}$ acceptors as $As:Ga$ is reduced from the 
As-rich condition. The maximum of $p$ along y corresponds to the position at 
which As$_{\mathrm{Ga}}$ are minimized, the stoichiometric condition. A further 
reduction in $As:Ga$ leads to a Ga-rich surface that limits Mn incorporation, 
and leads to a ten-fold increase in surface roughness marking a transition 
from 2D to 3D film morphology (Sec. \ref{Sec:structural}). The y-position of the maximum in $p$ 
matches in four separate non-rotated growths demonstrating the 
reproducibility of this technique. When the Mn density is reduced, the 
hole-density peak narrows on the As-rich side, indicating that samples with 
lower Mn-doping density are more sensitive to the As$_{\mathrm{Ga}}$ compensation 
effect. This follows if $As:Ga$ controls the density of hole 
compensating defects independently from the doping density of Mn, as 
expected for this doping regime.

We note that for the 0.75{\%} Mn doped wafer, when $As:Ga \sim 10$ the hole 
compensation is nearly complete, $p \sim 10^{16}$ cm$^{ - 3}$. For $As:Ga 
\sim 9$, the compensation is minimized, $p \sim  10^{19}$ cm$^{ - 3}$. 
In this case, a 10{\%} reduction in $As:Ga$ reduces the As$_{\mathrm{Ga}}$ 
compensation by three orders of magnitude. This demonstrates the strong 
sensitivity of the As$_{\mathrm{Ga}}$ incorporation to $As:Ga$, and the high degree of 
As-flux control, within 10{\%}, necessary to properly minimize this defect. 
Previous studies of the effect of $As:Ga$ ratio reported only two or three 
values of this parameter in increments of 200{\%} or 
more.\cite{Schott:2003,Sadowski:2004,Avrutin:2005}

In the Ga-rich position ($As:Ga < 9$), the hole density decreases along with 
$As:Ga$, but remains constant as the Mn doping density is varied, Fig. \ref{fig5}b. This 
behavior is explained by a solubility limit for Mn, which is below any of 
the Mn doping levels used here. Decreasing $As:Ga$ below the 
stoichiometric value leads to a Ga-rich surface that serves as a kinetic 
barrier for the incorporation of group-III substitutional dopants, such as 
Mn. One might consider the alternative possibility that interstitial 
incorporation of Mn might occur in these circumstances and could explain the 
reduction in hole density in the Ga-rich region. However, if Mn$_{\mathrm{i}}$ donors 
were incorporating and compensating the Mn$_{\mathrm{Ga}}$ acceptors, then the 
carrier density should change as the Mn-doping density is altered, which is 
not the case here. Preliminary SIMS data show a decreasing concentration of Mn in the Ga-rich region as the thickness increases, whereas the As-rich region reveals a relatively constant Mn depth profiles throughout the Mn-doped layer. In the Ga-rich region the Mn density can be $\sim 10^{20}$ cm$^{-3}$ lower than in the As-rich region.

The position of the hole density peak can be controlled by adjusting $As:Ga$ 
at y = 0, which shifts the position of the stoichiometric condition along y. 
To demonstrate this, two wafers with identical Mn concentration of 1.5 {\%} 
are grown but with different $As:Ga$ at y = 0. The variation of $p$ along y is 
plotted in Fig. \ref{fig5}c. By increasing $As:Ga$ at y = 0 from 8:1 to 9:1, the peak 
position (stoichiometric condition) is shifted by $\sim $ 1 cm toward the 
center of the wafer in a direction away from the As source. This shift is 
observable by eye as a change in the position of the transition region, 
where the surface changes from mirror finish to hazy. Converting the 
y-position to $As:Ga$ should fully overlap the hole density peaks, Fig. \ref{fig5}d. 
The peaks do not completely overlap indicating that additional factors must be considered, such as the variation in the growth rate across non-rotated wafers, which may play a role in the density of compensating As$_{\mathrm{Ga}}$ defects. In the present analysis we have considered only the effect of As:Ga.

Finally, we compare samples of equal Mn doping density grown with and without 
rotation. The hole densities of the rotated samples, Fig. \ref{fig1}b, match the hole 
densities for the non-rotated samples with similar Mn doping in the As-rich 
condition of $As:Ga \sim  10.5$, Fig. \ref{fig5}b. This indicates that the rotated 
sample set, presented in Fig. \ref{fig1}b, is grown under a slightly As-rich 
condition, on the right side of the hole density peak of Fig. \ref{fig5}b.

\subsection{Curie temperature gradient}
\label{Subsec:tcgradient}

\begin{figure}\includegraphics{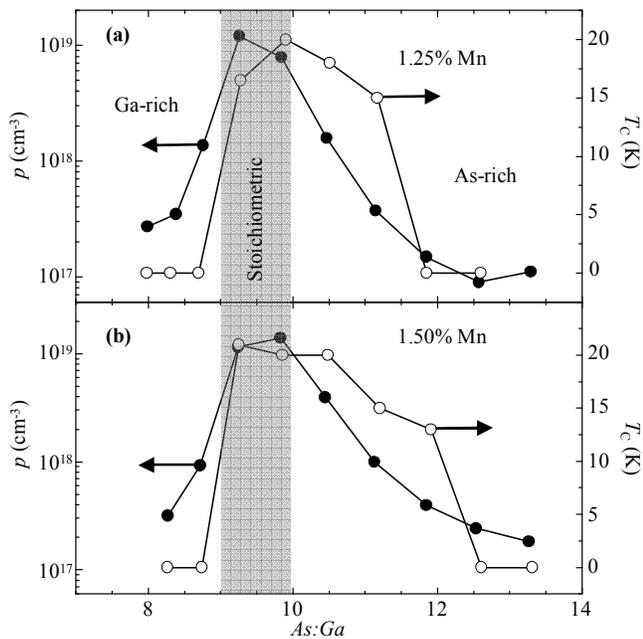}\caption{\label{fig6}
    The Curie temperature gradient in non-rotated (Ga,Mn)As films 
		due to As-compensation. The hole density ($p$) and $T_{\mathrm{C}}$ are plotted as a 
		function of $As:Ga$ for two wafers with different Mn concentrations, 
		(a) 1.25{\%}, and (b) 1.50{\%}. Lines guide the eye. The stoichiometric 
		region is shaded grey, where excess As is minimized.}\end{figure}

The Curie temperature variation with $As:Ga$ for two non-rotated (Ga,Mn)As films 
is plotted in Fig. \ref{fig6} a and b, together with the hole density variation.
$T_{\mathrm{C}}$ is determined from scans similar to those plotted in Fig. \ref{fig1}d. 
$T_{\mathrm{C}}$ tracks the $p$ variation with $As:Ga$, displaying a maximum in Curie 
temperature in the same region where the hole density peaks. In the As-rich 
portion of the wafer, As$_{\mathrm{Ga}}$ donors limit the hole density and therefore 
$T_{\mathrm{C}}$. This effect is very clear, for example, with $As:Ga > 12$, where 
ferromagnetism is completely suppressed in both 1.5{\%} and 1.25{\%} samples 
due to hole compensation from As non-stoichiometry in the form of As$_{\mathrm{Ga}}$ donors. 

\begin{figure}\includegraphics{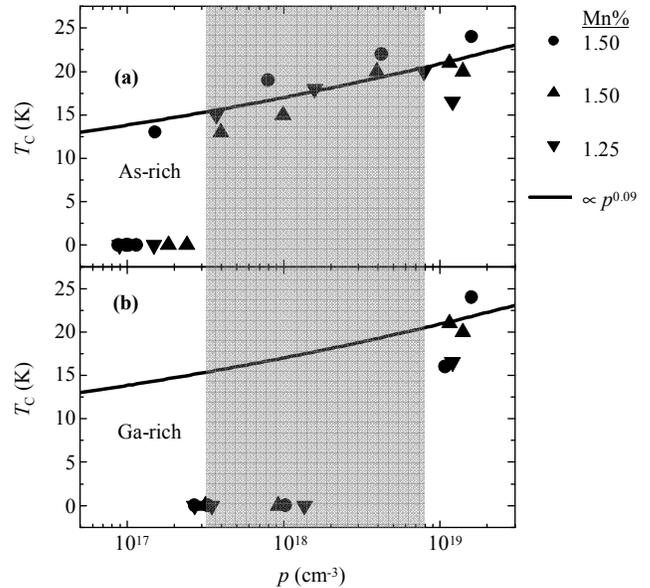}\caption{\label{fig7}
     Curie temperature ($T_{\mathrm{C}}$) versus hole density ($p$) in several 
		non-rotated (Ga,Mn)As films showing deviation from $p^{0.33}$. (a) Data are 
		plotted from the As-rich region of the wafer ($As:Ga \ge 9$), and (b) from 
		the Ga-rich region ($As:Ga \le 9$). The shaded region marks $p$-values at 
		which ferromagnetism is observed in As-rich material (a) and not in Ga-rich 
material (b). The line fit of data in (a) to $T_{\mathrm{C}} \propto p^{0.09}$ is replotted in (b).}\end{figure}

The width and shape of the $T_{\mathrm{C}}$ and hole density peaks are quite 
different. In general, the $T_{\mathrm{C}}$ peaks are wider than the hole density 
peaks on the As-rich side, but are sharper than the hole density peak on the 
Ga-rich side of the wafer. To explain this behavior, we replot data from 
several of the non-rotated (Ga,Mn)As wafers to examine the dependence of 
$T_{\mathrm{C}}$ on the hole density. This dependence is plotted for the As-rich 
region ($As:Ga \ge 9$) in Fig. \ref{fig7}a, and for the Ga-rich region ($As:Ga \le 9$) in Fig. \ref{fig7}b. 
		
In the As-rich region, ferromagnetism is observed over a broad range in hole 
density from 10$^{17}$ to 10$^{19}$ cm$^{ - 3}$. For Ga-rich material, 
however, ferromagnetism is not observed over the same large range in hole 
densities where ferromagnetism is observed in the As-rich material, shaded 
grey. As described in Sec. \ref{Sec:structural} and \ref{Subsec:pgradient}, hole density and film morphology measurements 
indicate that in the Ga-rich region Mn solubility is limited to well-below 
the level required for ferromagnetism, $<$ 0.5 {\%}. This is probably due to 
the kinetic disadvantage of a group-III dopant on a heavily Ga-rich surface 
because of out-competition. The discrepancy in Fig. \ref{fig7} between the 
Ga-rich region and As-rich region offers further support for this 
hypothesis. In this interpretation, Ga-rich conditions limit Mn$_{\mathrm{Ga}}$ to 
low doping densities of less than 10$^{19}$ cm$^{ - 3}$ ( $<$ 0.05{\%} Mn). 
Whereas for similar hole densities in the As-rich material Mn$_{\mathrm{Ga}} \sim  
10^{20}$ cm$^{ - 3}$ ($\sim $1{\%} Mn). Thus ferromagnetism is possible in 
the As-rich material since there are sufficient Mn$_{\mathrm{Ga}}$ acceptors, but the 
$T_{\mathrm{C}}$ is limited by hole compensation due to As$_{\mathrm{Ga}}$ donors. In the 
Ga-rich material, As$_{\mathrm{Ga}}$ compensating donors are not present, but the 
Mn$_{\mathrm{Ga}}$ density is well below the ferromagnetic limit.

The data in Fig. \ref{fig7}a for which $T_{\mathrm{C}} \ne $ 0 are fit to $T_{\mathrm{C}} \propto p^{0.09}$ (line). This fit describes the data from two non-rotated 
wafers with Mn = 1.5{\%} (Fig. \ref{fig5}c) and one wafer with 1.25{\%} Mn over a 
large range in $p$ and $T_{\mathrm{C}}$, over 13 data points. In the rotated growths 
presented in Fig. \ref{fig1}, $p^{0.33}$ behavior is observed for samples with $ \ge 
$ 1.5{\%} Mn. There is no discrepancy between the non-rotated and rotated 
(Ga,Mn)As films if we conjecture that a change in the exponent from 0.33 to 0.09 
occurs as the doping is reduced below 1.5{\%} Mn. Since these doping levels 
are close to the ferromagnetic limit, a change in the dependence of 
ferromagnetism is not unexpected. The impurity band is fully merged with the 
valence band in the metallic region of high Mn-doping, where the $p^{0.33}$ 
is observed. But at low Mn-doping levels close to the insulating regime, the 
impurity band is separated from the valence band (Mott gap) where a 
disorder-based model of ferromagnetism is probably more 
accurate.\cite{Yang:2003} Erwin and Petukhov predicted that a strong 
deviation from the metallic behavior of $T_{\mathrm{C}} \propto p^{0.33}$ 
should occur in the low Mn-doping regime, $\sim $1.5{\%} Mn, with $T_{\mathrm{C}}$ 
reaching a maximum in a partially compensated condition (lower $p$) and 
ferromagnetism disappearing for zero compensation (highest $p$); their model 
assumed that compensation was due to Mn$_{\mathrm{i}}$ alone.\cite{Erwin:2002} 
This behavior is not clearly seen in our data, however the observed 
As$_{\mathrm{Ga}}$ compensation was not treated in the model of Erwin and Petukhov.

\section{Conclusions}
\label{Sec:conclusions}

The electronic and magnetic properties of (Ga,Mn)As reveal a strong sensitivity 
to As$_{\mathrm{Ga}}$ donors that compensate Mn$_{\mathrm{Ga}}$ acceptors. By varying
$As:Ga$, the hole densities of (Ga,Mn)As films are varied by more than two 
orders of magnitude, and, as a result, ferromagnetism can be enhanced or completely 
suppressed. For As-rich growth, films show the maximum 
smoothness, but lowest $p$ and $T_{\mathrm{C}}$. If $As:Ga$ is reduced, the film 
roughness increases up to the stoichiometric condition, where $p$ and $T_{\mathrm{C}}$ 
are maximized. For Ga-rich growth, the film roughness increases by an order 
of magnitude as excess Ga accumulates on the surface. The group-III rich 
surface suppresses substitutional Mn incorporation leading to lower $p$ and 
T$_{C }$ values than in the As-rich condition. In the low Mn-doped regime, 
where As-compensation limits hole density and ferromagnetic ordering, a 
strong deviation from the usual $T_{\mathrm{C}} \propto p^{0.33}$ behavior is 
observed with $T_{\mathrm{C}} \propto p^{0.09}$ for x $<$ 1.5{\%}. This 
dependence applies equally well to stoichiometric (Ga,Mn)As as it does to 
heavily As$_{\mathrm{Ga}}$ compensated (Ga,Mn)As. This change in the hole density 
dependence of $T_{\mathrm{C}}$ occurs in the doping region where the effects of a 
Mott gap and disorder are expected to alter the mechanism of 
ferromagnetism.\cite{Erwin:2002,Yang:2003} Our results 
emphasize that precise control of hole compensation due to As$_{\mathrm{Ga }}$ is 
critical to maximizing the value of $T_{\mathrm{C}}$'s in the low Mn-doping regime. The 
combinatorial MBE growth approach used here could be applied to more complex 
LT grown GaAs-based heterostructures in which $As:Ga$ is a strong control 
parameter. The rapid and high-resolution mapping of growth parameter 
phase-space by this technique allows for systematic investigations not 
practicable using more traditional methods.

\begin{acknowledgments}
The authors thank D. W. Steuerman for SEM assistance, M. Poggio for helpful 
criticism, and J. H. English for MBE technical know-how. This work was 
financially supported by DARPA, ONR, and the NSF. We made use of MRL Central Facilities supported by the MRSEC Program of the NSF (DMR05-20415).
\end{acknowledgments}


\begin{thebibliography}{24}
\bibitem{Ohno:1998} H. Ohno, Science \textbf{281}, 951 (1998).
\bibitem{Dietl:2000} T. Dietl, H. Ohno, F. Matsukura, J. Cibert, and D. Ferrand, Science \textbf{287}, 1019 (2000).
\bibitem{Jungwirth:2006} T. Jungwirth, J. Sinova, J. Masek, J. Kucera, and A. H. Macdonald, cond-mat/0603380.
\bibitem{Macdonald:2005} A. H. Macdonald, P. Schiffer, and N. Samarth, Nature Materials \textbf{4}, 195 (2005).
\bibitem{Erwin:2002} S. C. Erwin and A. G. Petukhov, Phys. Rev. Lett. \textbf{89}, 227201 (2002).
\bibitem{Edmonds:2004} K. W. Edmonds, P. Boguslawski, K. Y. Wang, R. P. Campion, S. N. Novikov, N. R. S. Farley, B. L. Gallagher, C. T. Foxon, M. Sawicki, T. Dietl, M. B. Nardelli, and J. Bernholc, Phys. Rev. Lett. \textbf{92}, 037201 (2004).
\bibitem{Ku:2003} K. C. Ku, S. J. Potashnik, R. F. Wang, S. H. Chun, P. Schiffer, N. Samarth, M. J. Seong, A. Mascarenhas, E. Johnston-Halperin, R. C. Myers, A. C. Gossard, and D. D. Awschalom, Appl. Phys. Lett. \textbf{82}, 2302 (2003).
\bibitem{Liu:1995} X. Liu, A. Prasad, J. Nishio, E. R. Weber, Z. Liliental-Weber, and W. Walukiewicz, Appl. Phys. Lett. \textbf{67}, 279 (1995).
\bibitem{Missous:1994} M. Missous and S. O'Hagan, J. Appl. Phys. \textbf{75}, 3396 (1994).
\bibitem{Bliss:1992} D. E. Bliss, W. Walukiewicz, J. W. Ager Iii, E. E. Haller, K. T. Chan, and S. Tanigawa, J. Appl. Phys. \textbf{71}, 1699 (1992).
\bibitem{Boeck:1996} J. De Boeck, R. Oesterholt, A. Van Esch, H. Bender, C. Bruynseraede, C. Van Hoof, and G. Borghs, Appl. Phys. Lett. \textbf{68}, 2744 (1996).
\bibitem{Yang:2003} S. R. E. Yang and A. H. MacDonald, Phys. Rev. B \textbf{67}, 155202 (2003).
\bibitem{Schott:2001} G. M. Schott, W. Faschinger, and L. W. Molenkamp, Appl. Phys. Lett. \textbf{79}, 1807 (2001).
\bibitem{Ohno:1996} H. Ohno, A. Shen, F. Matsukura, A. Oiwa, A. Endo, S. Katsumoto, and Y. Iye, Appl. Phys. Lett. \textbf{69}, 363 (1996).
\bibitem{Schott:2003} G. M. Schott, G. Schmidt, G. Karczewski, L. W. Molenkamp, R. Jakiela, A. Barcz, and G. Karczewski, Appl. Phys. Lett. \textbf{82}, 4678 (2003).
\bibitem{Sadowski:2004} J. Sadowski and J. Z. Domagala, Phys. Rev. B \textbf{69}, 075206 (2004).
\bibitem{Campion:2003} R. P. Campion, K. W. Edmonds, L. X. Zhao, K. Y. Wang, C. T. Foxon, B. L. Gallagher, and C. R. Staddon, J. Crys. Grow. \textbf{247}, 42 (2003).
\bibitem{Foxon:2005} C. T. Foxon, R. P. Campion, K. W. Edmonds, L. Zhao, K. Wang, N. R. S. Farley, C. R. Staddon, and B. L. Gallagher, J. Mater. Sci. - Mater. Elec. \textbf{15}, 727 (2005).
\bibitem{Avrutin:2005} V. Avrutin, D. Humienik, S. Frank, A. Koeder, W. Schoch, W. Limmer, R. Sauer, and A. Waag, J. Appl. Phys. \textbf{98}, 023909 (2005).
\bibitem{Schippan:2000} F. Schippan, M. Kastner, L. Daweritz, and K. H. Ploog, Appl. Phys. Lett. \textbf{76}, 834 (2000).
\bibitem{Farrow:1995} R. A. Kubiak, S. M. Newstead, and P. Sullivan, in \textit{Molecular Beam Epitaxy: Applications to Key Materials}, edited by R. F. C. Farrow, (Noyes, 1995).
\bibitem{Jackson:1} A. W. Jackson, R. C. Myers, and A. C. Gossard, (unpublished).
\bibitem{Poggio:2005} M. Poggio, R. C. Myers, N. P. Stern, A. C. Gossard, and D. D. Awschalom, Phys. Rev. B \textbf{72}, 235313 (2005).
\bibitem{Brandt:2002} O. Brandt, P. Waltereit, and K. H. Ploog, J. Phys. D - Appl. Phys. \textbf{35}, 577 (2002). ``MadMax'' executable program courtesy of Patrick Waltereit of Fraunhofer IAF.
\end{thebibliography}
\end{document}